\documentclass{article}
\usepackage[nonatbib, final]{nips_2017}
\usepackage[utf8]{inputenc} 
\usepackage[T1]{fontenc}    
\usepackage{hyperref}       
\usepackage{url}            
\usepackage{booktabs}       
\usepackage{amsfonts}       
\usepackage{nicefrac}       
\usepackage{microtype}      
\usepackage{graphicx}
\usepackage{amsmath}
\usepackage{bm}
\usepackage{algorithmicx}
\usepackage{algorithm}
\usepackage{algpseudocode}
\usepackage[table]{xcolor}

\title{Cross-Attribute Matrix Factorization Model with Shared User Embedding}

\author{%
  Wen Liang \\
  University of California, San Diego \\
  La Jolla, CA 92093 \\
  \texttt{wel245@eng.ucsd.com} \\
  \And
  Zeng Fan \\
  University of California, San Diego \\
  La Jolla, CA 92093 \\
  \texttt{zengfan1109@gmail.com} \\
  \And
  Youzhi Liang \\
  Department of Computer Science \\
  Stanford University \\
  Stanford, CA 94305 \\
  \texttt{youzhil@stanford.edu} \\
  \And
  Jianguo Jia \\
  Department of Computing \\
  Hong Kong Polytechnic University\\
  Hong Kong, China \\
  \texttt{jianguo1.jia@connect.polyu.hk}
}

\begin{document}

\maketitle

\begin{abstract}

Over the past few years, deep learning has firmly established its prowess across various domains, including computer vision, speech recognition, and natural language processing. Motivated by its outstanding success, researchers have been directing their efforts towards applying deep learning techniques to recommender systems. Neural collaborative filtering (NCF) and Neural Matrix Factorization (NeuMF) refreshes the traditional inner product in matrix factorization with a neural architecture capable of learning complex and data-driven functions. While these models effectively capture user-item interactions, they overlook the specific attributes of both users and items. This can lead to robustness issues, especially for items and users that belong to the "long tail". Such challenges are commonly recognized in recommender systems as a part of the cold-start problem. A direct and intuitive approach to address this issue is by leveraging the features and attributes of the items and users themselves. In this paper, we introduce a refined NeuMF model that considers not only the interaction between users and items, but also acrossing associated attributes. Moreover, our proposed architecture features a shared user embedding, seamlessly integrating with user embeddings to imporve the robustness and effectively address the cold-start problem. Rigorous experiments on both the Movielens and Pinterest datasets demonstrate the superiority of our Cross-Attribute Matrix Factoriztion model, particularly in scenarios characterized by higher dataset sparsity.

\end{abstract}

\section{Introduction}

Recommender systems have become an indispensable component of many online platforms, guiding users through a plethora of content to find what best matches their preferences and needs. Traditional methods, such as collaborative filtering \cite{CF} and matrix factorization \cite{MF}, have been successful in capturing the latent preferences of users based on historical interactions~\cite{resnick1997recommender}. The rise of deep learning has revolutionized numerous domains, including computer vision where it has enabled breakthroughs in image recognition \cite{resnet, imagenet},  natural language processing \cite{transformer, bert}, and speech recognition \cite{speech}. Variational techniques and hybrid architectures of deep learning with data augmentations \cite{liang2023miamix} elevate the performance of modeling to a higher standard~\cite{liang2023reswcae, liang2023structural}. Leveraging these advancements, models such as NeuMF and NCF have incorporated neural networks in the realm of recommender systems, advancing beyond the linearity of matrix factorization to allow for more intricate representations of user-item interactions \cite{He}. However, a gap remains in considering the intrinsic attributes of users and items, which can be pivotal in addressing challenges such as the cold-start problem and ensuring robustness, especially for long-tail items and users. 

While the NeuMF and NCF model mark significant advancements in the field, they primarily focus on the interactions derived from user-item matrices without tapping into the wealth of information contained within intrinsic attributes of users and items. By overlooking these attributes, the models potentially miss out on crucial signals that could enhance recommendation quality, especially in scenarios where interaction data is sparse or nonexistent, such as in the cold-start problem. To address this, the Attribute-aware Deep CF model incorporates item and user features into the matrix factorization (MF) model \cite{pinterest}. It transforms these features into learnable embeddings, subsequently integrating them into the multi-layer perceptron (MLP). However, we believe there is potential to further enhance this approach. Specifically, there's an opportunity to exploit more features by cross-referencing user attributes with item attributes and vice versa.

Our \textit{Cross-Attribute Matrix Factorization Model with Shared User Embedding} offers several distinctive contributions to the realm of recommender systems:

\begin{enumerate}
    \item \textbf{Incorporation of a Shared User Embedding}: One of the major contributions of our model is the introduction of a shared user embedding. In situations where we encounter cold-start users, this shared embedding facilitates a basic recommendation process. It mitigate the robustness problems posed by relying on either randomly-initialized embeddings or immaturely learned user embeddings, which often aren't reliable for providing meaningful recommendations.
    
    \item \textbf{Enhancement of the Existing MF Model with Cross-Attribute Interactions}: Building on the foundation of the current matrix factorization (MF) model, we've added cross-attribute matrix factorization capabilities. To elaborate, each user embedding interacts with every item attribute embedding, and vice versa; every item interacts with user attribute embeddings. This intricate design ensures that we harness the full potential of existing data and features, fostering richer and more insightful recommendation outcomes.
\end{enumerate}

Our model are further validated through a series of experiments. We used two benchmark datasets - Movielens and Pinterest \cite{pinterest} - consistently underscore the enhanced performance of our model. Notably, CAMF shines with pronounced superiority in environments marked by heightened dataset sparsity, thereby showcasing its practical applicability in real-world settings where user-item interactions may be limited.

\section{Related Works}

\subsection{Matrix Factorization}

\begin{figure}
  \centering
  \includegraphics[width=0.4\textwidth]{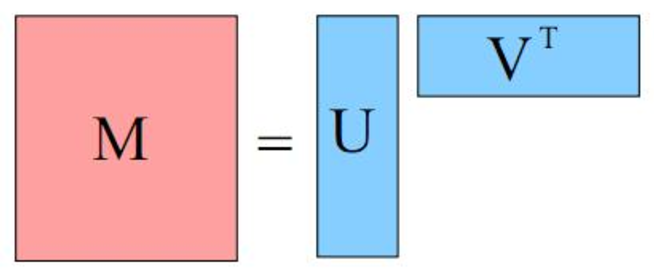}
  \caption{Matrix Factorization}
  \label{fig:MF} 
\end{figure}

MF convert each user and item to a real-valued vector of latent variables. The MF estimates the interaction between user and item by using the inner product of two vectors:
\begin{equation}
	r(u, i) = \gamma_u \cdot \gamma_i,
\end{equation}
where the $\gamma_u$ and $\gamma_i$ denotes the latent factor for user $u$ and item $i$, the $r(u,i)$ denotes the interaction between this user and item. This method is very popular by combining good scalability with predictive accuracy. In addition, they offer much flexibility for modeling
various real-life situations. However, the matrix factorization has several limitations that it is prone to recommend popular items and is vulnerable to sparse dataset.

\subsection{NCF, GMF and NeuMF Model}

\begin{figure}
  \centering
  \includegraphics[width=0.8\textwidth]{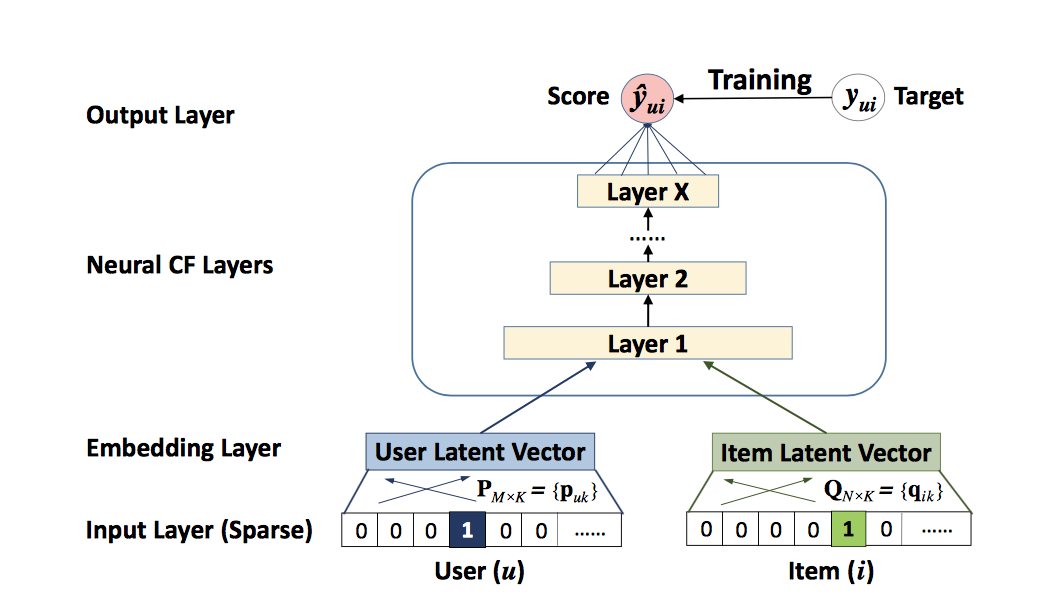}
  \caption{Neural collaborative filtering framework}
  \label{fig:NCF} 
\end{figure}

Neural Collaborative Filtering (NCF)~\cite{He} serves as a foundational framework to model the interaction between users and items using neural networks. The general frame work of NCF is shown in Figure~\ref{fig:NCF}. Within the NCF paradigm, two primary models emerge: GMF (Generalized Matrix Factorization) and MLP (Multi-Layer Perceptron). GMF can be seen as a neural analog to traditional matrix factorization techniques, capturing linear patterns in user-item interactions. On the other hand, MLP is designed to discover intricate non-linear patterns. The NeuMF model blends the linear capabilities of GMF with the rich expressiveness of MLP to further improve recommendation performance. We mainly focus on GMF model and NeuMF model in our work. The mapping function of the GMF layer is defined as following:

\begin{equation}
	\mathbf{\phi}(p_u,q_i) = p_u \odot q_i\end{equation}
where $\odot$ denotes the element-wise product of vectors. Then this vector can be projected to the output layer:
\begin{equation}
	\hat{y}_{ui} = \mathbf{a_{out}}(h^T (p_u \odot q_i)).
\end{equation}

\begin{figure}
  \centering
  \includegraphics[width=0.8\textwidth]{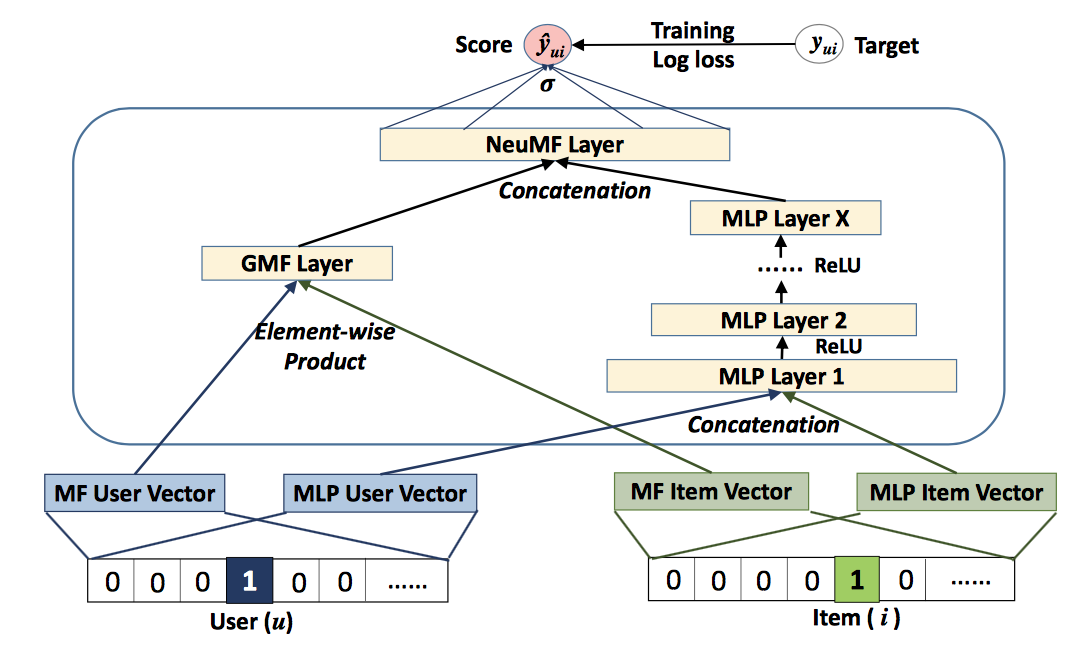}
  \caption{Neural Matrix Factorization Model}
  \label{fig:NeuMF} 
\end{figure}

In the work presented by He et al.~\cite{He}, the GMF (Generalized Matrix Factorization) layer, characterized by its activation function \(a_{\text{out}}\) and edge weights \(h\), offers a pathway to revert back to the traditional matrix factorization model. As depicted in Figure~\ref{fig:NeuMF}, the NeuMF (Neural Matrix Factorization) model is an integration of both the GMF and the MLP (Multi-Layer Perceptron) models. Within this architecture, the GMF and MLP layers independently process the input embeddings. Subsequently, their outputs are concatenated, forming the basis for the NeuMF layer.

\subsection{Attributed-Aware Deep Collaborative Filtering Model}

\begin{figure}
  \centering
  \includegraphics[width=0.8\textwidth]{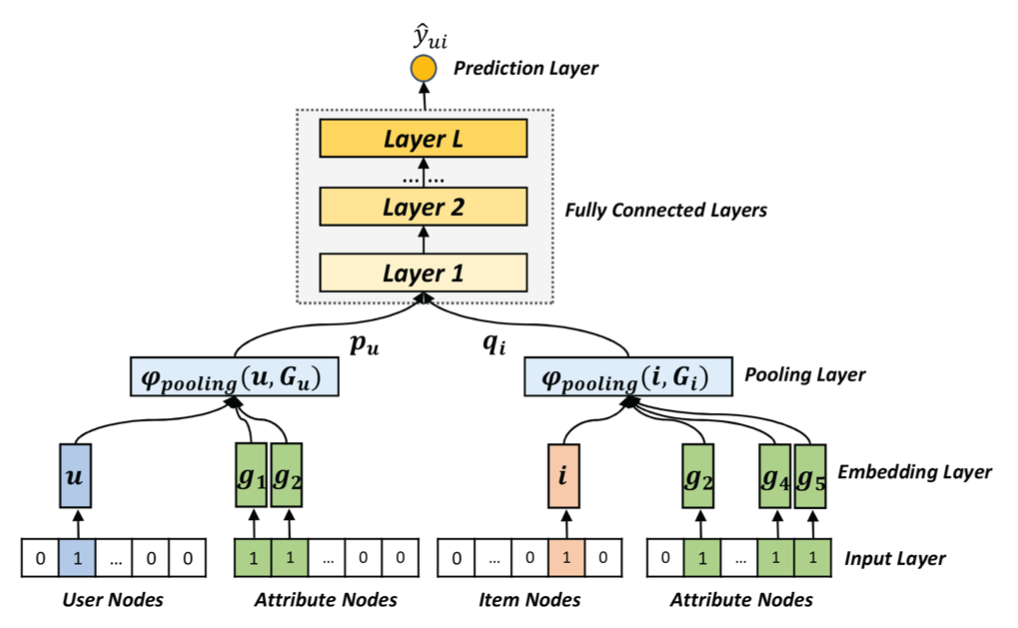}
  \caption{Attribute-aware Neural CF Model}
  \label{fig:ADCF} 
\end{figure}

The NCF framework takes into account the user-item interaction information, but it neglects the intrinsic attribute information of users/items. Wang et al.~\cite{Wang} proposed an Attribute-aware deep CF model, as depicted in Figure~\ref{fig:ADCF}, which incorporates the attributes of users/items. They introduced a pairwise pooling layer following the embedding layer to effectively capture the relationship between users/items and their corresponding attributes. The operation for pairwise pooling is defined as:

\begin{equation}
	p_u  = \mathbf{\varphi_{pairwise}}(u,{g^t_u}) = \sum \limits_{t = 1}^{V_u} u\odot g^t_u + \sum \limits_{t = 1}^{V_u} \sum \limits_{t\prime = t+1}^{V_u} g^t_u\odot g^t\prime_u
\end{equation}
\begin{equation}
	q_i  = \mathbf{\varphi_{pairwise}}(i,{g^t_i}) = \sum \limits_{t = 1}^{V_i} i\odot g^t_i + \sum \limits_{t = 1}^{V_i} \sum \limits_{t\prime = t+1}^{V_i} g^t_i\odot g^t\prime_i
\end{equation}
After calculate $p_u$ and $q_i$, the model takes the element-wise product of $p_u$ and $q_i$ into MLP layers for the final prediction.

\section{Method}

\subsection{Shared User Embedding}

Our experiments with GMF, NeuMF, and Attribute-aware Deep CF models showed that the GMF model trained faster and had nearly the same performance as the more complex NeuMF. The detailed results are discussed later. Due to its efficiency and effectiveness, we chose to focus on the GMF model.

Our goal was to improve recommendations in scenarios with sparse data and for new users or items. To address the cold-start and sparsity challenges, we use a shared user embedding embedding might encapsulate overall user traits. As an illustrative case, encountering a novel user or item could harness this shared embedding, furnishing more generic and resilient recommendation outcomes. To combine this shared user embedding with a specified user embedding, we introduce a weight $\alpha$ ranging from 0 to 1 to adjust the balance between them by controlling the activation of the specified user embedding:

\begin{equation}
u_{merged} = \alpha u_{shared} + (1 - \alpha) u_{embedded}
\end{equation}
\begin{equation}
\mathbf{\phi}(u_{merged},i) = u_{merged} \odot i
\end{equation}

In alignment with the GMF paradigm delineated earlier, the weighted user vector undergoes an element-wise multiplication with the item vector. The balancing weight, $\alpha$, is determined by a singular hidden layer, taking another ensemble of item and user attributes as inputs:

\begin{equation}
\mathbf{z} = 
\begin{bmatrix}
u\\
u_{\text{attr}} \\
i \\
i_{\text{attr}}
\end{bmatrix}
\end{equation}

\begin{equation}
\alpha = \text{Sigmoid}(\mathbf{w}^T \mathbf{z} + \mathbf{b})
\end{equation}

Herein, $\mathbf{z}$ symbolizes the layer input, which incorporates attributes like item popularity and user historical interaction frequency. For instance, if attributes indicate an new user, the model might assign greater emphasis on the shared user vector, recommending universally acclaimed movies. Conversely, for universally appealing movies, the model could diminish the weight on specifed user embedding, offering recommendations less tailored to individual tastes.

\subsection{Cross-Attribute Matrix Factorization Model}

Another pivotal enhancement in our approach lies in the explicit incorporation of attribute information. Specifically, we multiplied the user vectors with item attributes and item vectors with user attributes. To accommodate this, we introduced dedicated embeddings tailored for attributes. This setup is not just a mere inclusion of attributes; it fundamentally changes the representation learning. By allowing these interactions, we aim to learn the underlying patterns and relationships between users, items, and their inherent attributes.

After facilitating these interactions, the resultant vectors are concatenated. This consolidated representation captures both the standalone and cross-attribute information. To process this rich representation further, we direct it through several neural network layers. These layers are designed to abstract higher-level patterns from the combined input.

\begin{figure*}
 \center
  \includegraphics[width=0.8\textwidth]{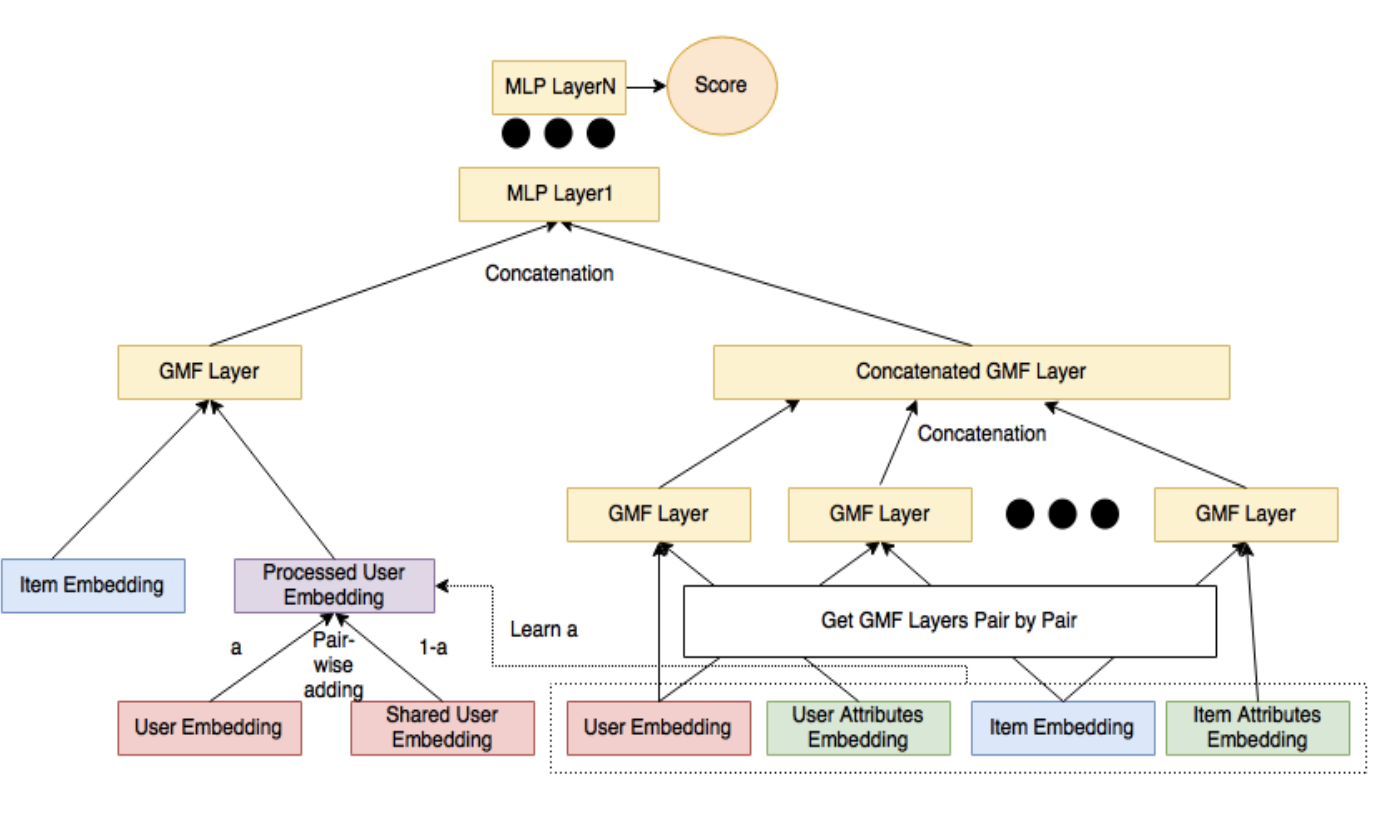}
  \caption{Proposed Attribute-aware Model}
  \label{fig:Our_Model}
\end{figure*}

Our comprehensive model, which encompasses the previously discussed Shared User Embedding and the current Cross-Attribute interactions, is illustrated in Figure~\ref{fig:Our_Model}.

\section{Experiments}

In our evaluations, we leverage two prominent datasets: MovieLens and Pinterest. Both have been previously used in \cite{He, pinterest} due to their rich attributes and feasibility for preprocessing.

\subsection{MovieLens Dataset}

For MovieLens, we employ the version containing 1 million ratings, ensuring that each user has rated at least 20 movies. Explicit ratings are transformed: a rating from a user denotes a label of 1 for the movie. To supplement this, we randomly sample 99 unobserved entries, tagging them as negative samples with a label of 0.

The dataset also provides user metadata encompassing gender, age, and occupation. This metadata is used as the user's attributes, denoted as \(g^t_u\), both in the Attribute-aware Deep CF model and our proposed architecture.

Movie data primarily comprise genres. Recognizing that movies can span multiple genres, each genre is treated as a distinct attribute. These are used as \(g^t_i\) during our experiments.

\subsection{Pinterest Dataset}

The Pinterest dataset, characterized by its enormity and sparsity, presents a unique challenge. With over 20\% of the users having a singular pin, assessing algorithmic efficacy becomes intricate. Thus, we refined the dataset to include only users with at least 10 pins, which are labeled as 1. Like the MovieLens dataset, 99 unobserved entries are sampled as negatives labeled as 0.

User data in this dataset provide insights into pins and page categories. We group every 40 pins together, such that a user with, say, 35 pins falls into the first group and a user with 41 pins falls into the second. The myriad of page categories, initially totaling 468, are consolidated into 45 main categories, given the observation that several categories possess scant samples. This structured data is then taken as the user's attribute input.

Items, represented by images, lack explicit attributes. Notably, an image may be pinned by diverse users. To derive an item's 'pin attribute', we employ the formula:
\begin{equation}
	Pins_{i}  = \sum \limits_{ (u\prime,i)\subset D }^{V_u} Pins_{u\prime i}
\end{equation}
In the equation, \(V_u\) signifies the user set in the dataset, and \(D\) encapsulates the user-item pairs. \(Pins_u\) stands for the user's pin count. This weighted summation offers insights into an item's popularity beyond a mere pin count. Analogous to user pins, items are grouped every 50, and this processed data serves as the item's attribute input.

\subsection{Evaluation and Metrics}

We use the leave-one-out evaluation in our experiment. For each user, we randomly choose one of user's interaction and the 99 negative samples mentioned in Data Preprocessing section as test set. We rank the test item among the 100 items. The performance of the ranked list is evaluated by first 10 Hit Ratio (HR@10) and Normalized Discounted Cumulative Gain (NDCG@10). In other word, HR measures whether the positive test item is within the top-10 ranked list, and the NDCG will assign higher scores to hits at top ranks. These 2 metrics are calculated for each user and the average score is reported.

\subsection{Parameters Setting}

We don't use the pretrained model from the paper. For each model, we initialize model parameters with a Gaussian distribution (mean = 0, standard deviation = 0.01). The models are trained with Adam optimizer with the batch size of 256. For stacked MLP layers, we use [32,16,8] as layer dimensions. During training process, we will randomly generate 4 negatives instance for each positive instance. He et al. term the last hidden layer of NCF as predictive factors. We evaluate the factors of [8,16,32].  

\subsection{Results}

\begin{figure}
  \centering
  \includegraphics[width=0.7\textwidth]{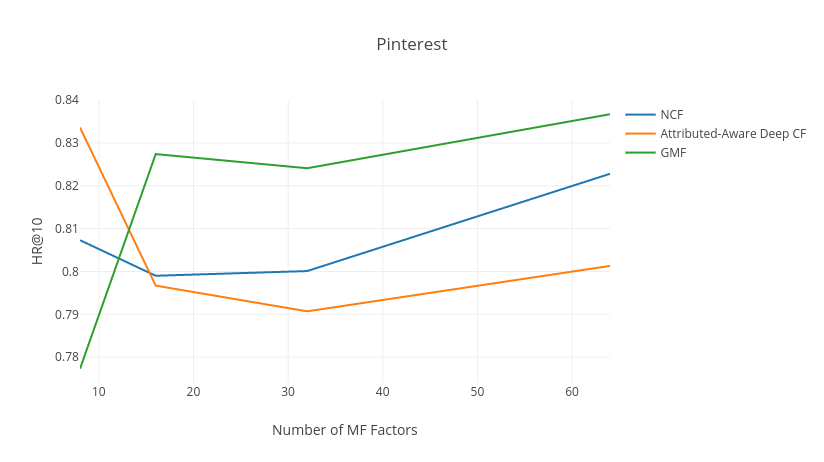}
  \caption{Performance of HR@10 on Pinterest}
  \label{fig:Pin_HR} 
\end{figure}

\begin{figure}
  \centering
  \includegraphics[width=0.7\textwidth]{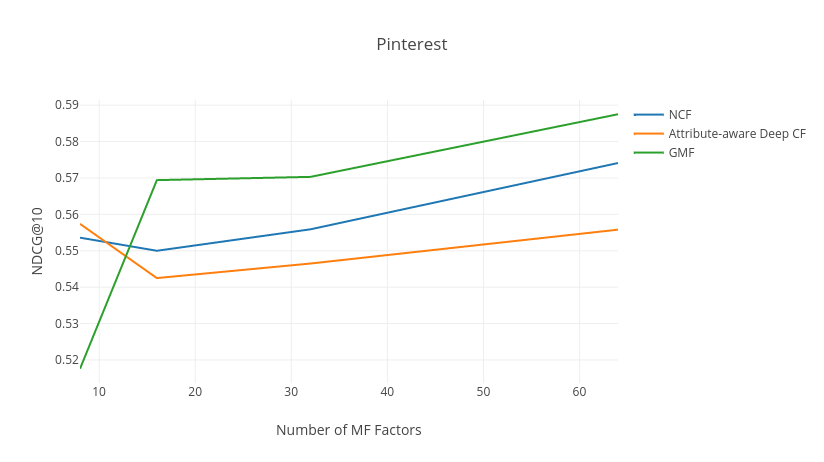}
  \caption{Performance of NDCG@10 on Pinterest}
  \label{fig:Pin_NDCG} 
\end{figure}

The following Figure~\ref{fig:ML_HR}, Figure~\ref{fig:ML_NDCG}, Figure~\ref{fig:Pin_HR} and Figure~\ref{fig:Pin_NDCG} show the performance of HR@10 and NDCG@10 with respect to the number of matrix factorization factors.

\begin{figure}
  \centering
  \includegraphics[width=0.7\textwidth]{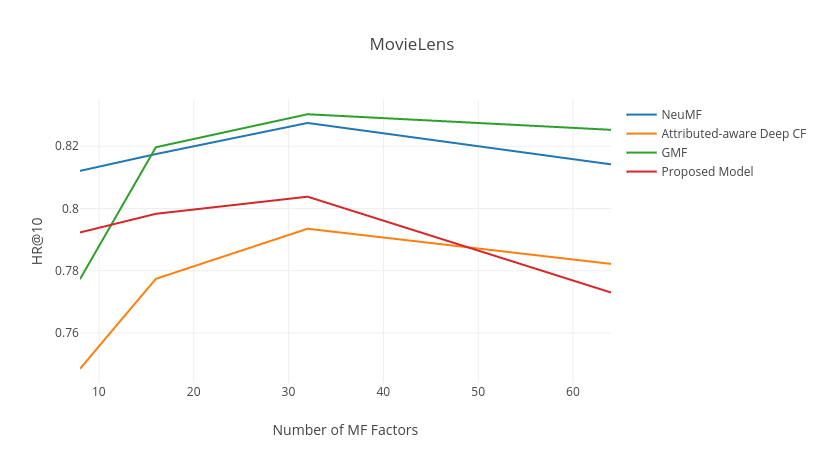}
  \caption{Performance of HR@10 on MovieLens}
  \label{fig:ML_HR} 
\end{figure}

\begin{figure}
  \centering
  \includegraphics[width=0.7\textwidth]{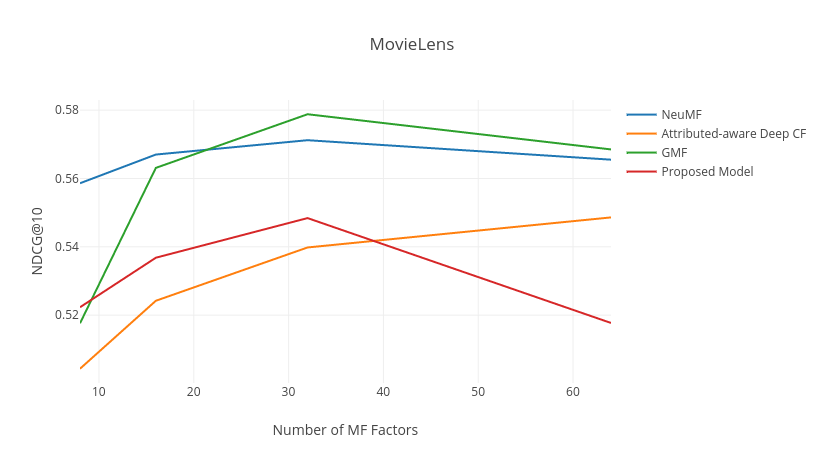}
  \caption{Performance of NDCG@10 on MovieLens}
  \label{fig:ML_NDCG} 
\end{figure}

\begin{figure}
  \centering
  \includegraphics[width=0.7\textwidth]{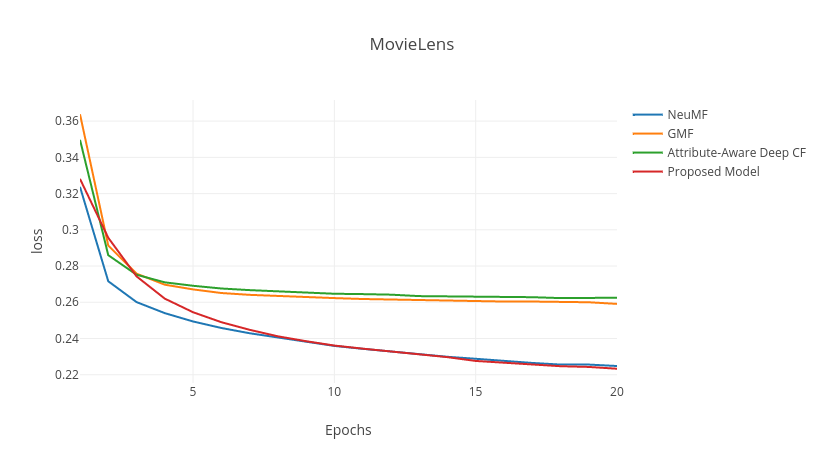}
  \caption{Training loss of all the models}
  \label{fig:loss} 
\end{figure}

We optimize all the models with log loss. The Figure~\ref{fig:loss} shows the training loss vs. epochs on MovieLens dataset for MLP, GMF, NeuMF, Attribute-aware Deep CF, and our proposed model.

The following tabel\ref{tab:ML_HR}, tabel\ref{tab:ML_NDCG}, tabel\ref{tab:Pin_HR} and tabel \ref{tab:Pin_NDCG} show the performance of HR@10 and NDCG@10 with respect to the number of matrix factorization factors.

\begin{table}[H]
\caption{Movielens HR@10 Results} 
\centering 
\begin{tabular}{c c c c c} 
\hline\hline 
\#Factors & NeuCF & GMF & AA Deep CF & Our Model\\ [0.5ex] 
\hline 
8 & 0.8121 & 0.7485 & 0.7773 & \textbf{0.7923}\\ 
16 & 0.8175 & 0.7774 & 0.8197 & \textbf{0.8213} \\
32 & 0.8275 & 0.7935 & 0.8303 & \textbf{0.8304} \\
\hline 
\end{tabular}
\label{tab:ML_HR} 
\end{table}

\begin{table}[H]
\caption{Movielens NDCG@10 Results} 
\centering 
\begin{tabular}{c c c c c} 
\hline\hline 
\#Factors & NeuCF & GMF & AA Deep CF & Our Model\\ [0.5ex] 
\hline 
8	& 0.5286	& 0.5043	& 0.5176	& \textbf{0.5308} \\
16	& 0.567	& 0.5242	& 0.5631	& 0.5668 \\
32	& 0.5712	& 0.5398	& 0.5788	& 0.5784 \\
\hline 
\end{tabular}
\label{tab:ML_NDCG} 
\end{table}

\begin{table}[H]
\caption{Pinterest HR@10 Results} 
\centering 
\begin{tabular}{c c c c c} 
\hline\hline 
\#Factors & NeuCF & GMF & AA Deep CF & Our Model\\ [0.5ex] 
\hline 
8 & 0.8073	& 0.8336 & 0.7773 & \textbf{0.8311} \\
16 & 0.799	& 0.7967 & 0.8274 & \textbf{0.8323} \\
32 & 0.8001	& 0.7907 & 0.8241 & \textbf{0.8610} \\
\hline 
\end{tabular}
\label{tab:Pin_HR} 
\end{table}

\begin{table}[H]
\caption{Pinterest NDCG@10 Results} 
\centering 
\begin{tabular}{c c c c c} 
\hline\hline 
\#Factors & NeuCF & GMF & AA Deep CF & Our Model\\ [0.5ex] 
\hline 
8	& 0.5536	& 0.5574	& 0.5176	& 0.5507 \\
16	& 0.55	& 0.5425	& 0.5694	& 0.5665 \\
32	& 0.5559	& 0.5465	& 0.5703	& \textbf{0.5920} \\
\hline 
\end{tabular}
\label{tab:Pin_NDCG} 
\end{table}

\subsection{Discussion}

The experimental results strongly underscore the robustness of our proposed model in comparison to existing models. Considering both the MovieLens and Pinterest datasets, our model consistently outperforms or is highly competitive with other methods across the board.

From the tables \ref{tab:ML_HR}, \ref{tab:ML_NDCG}, \ref{tab:Pin_HR}, and \ref{tab:Pin_NDCG}, it is evident that our model exhibits superior performance, especially with an increase in the number of matrix factorization factors. This superiority in HR@10 metrics is particularly significant given that our model is directly designed to tap into the rich information available in user-item interactions, without compromising on the intrinsic attributes.

The Attribute-Aware Deep CF (AA Deep CF) model's underwhelming performance on these recommendation tasks came as a surprise. One reason for its subpar results might be the pooling layer they adopted. By leveraging user, item, and attribute information, the model may inadvertently discard pivotal details, leading to a dilution of recommendation quality. Additionally, it's noteworthy to mention that the AA Deep CF model was primarily fashioned for social networks and travel recommendation tasks. While it might excel in scenarios with abundant user and item attributes, it evidently struggles when confronted with the data constraints of our datasets. The confluence of the AA Deep CF model with another model optimized for social network relations hints at design decisions which may not necessarily align with our task, and thus, the observed performance disparities.

In summary, our results illuminate the importance of crafting recommendation models tailored to the nuances of the dataset in question. Our model, by astutely leveraging both interaction data and inherent attributes, solidifies its place as a formidable contender in the recommendation system arena.

\section{Conclusion}

This research highlights two pivotal advancements in recommendation systems. First, we introduced the concept of a \textit{Shared User Embedding}, which offers a robust solution to the cold-start problem by reducing dependence on unstable embeddings. Concurrently, our refined matrix factorization approach, utilizing \textit{Cross-Attribute Interactions}, ensures an in-depth comprehension of user-item dynamics, leveraging every available attribute. Collectively, these innovations represent a substantial progression in providing nuanced and dependable recommendations, indicating an optimistic trajectory for subsequent research in this field.


\bibliographystyle{unsrt}
\bibliography{reference}
\end{document}